# Adiabatic cross-polarization

Jae-Seung Lee and A. K. Khitrin

Department of Chemistry, Kent State University, Kent OH 44242-0001, USA

**Abstract**

A scheme of adiabatic cross-polarization is described. It is based on demagnetization – remagnetization, when the Zeeman order of abundant nuclei in the laboratory frame is first adiabatically converted into the dipolar order, and then, into the Zeeman order of rare nuclei. The scheme, implemented with two low-power frequency-sweeping pulses, is very efficient for static samples and can significantly increase polarization of rare nuclei, compared to the conventional Hartmann-Hahn cross-polarization. The experimental examples are presented for solids, liquid crystal, and molecules in a liquid-crystalline solvent.

# INTRODUCTION

Cross-polarization (CP)[1,2] can be viewed as a process of transferring heat or, alternatively, order, between the two Zeeman reservoirs, representing two types of spins. In solid-state NMR, dipole-dipole interaction between heteronuclei is reduced to zz-terms only and, therefore, it is convenient to realize CP in the rotating frame. Strong resonance radio-frequency (RF) fields on both nuclei, with matched nutation frequencies (Hartmann-Hahn condition), convert this zz-interaction, in doubly rotating frame, into time-independent zero-order (flip-flop) average Hamiltonian, responsible for the thermal contact between two Zeeman reservoirs. This contact gradually equalizes Zeeman spin temperatures in the rotating frame. When the RF fields are matched, i.e. their amplitudes in frequency units are the same, equal spin temperatures mean equal polarizations for the two types of spins. In the limit of small concentration of rare nuclei, the change of spin temperature and polarization of the abundant nuclei is negligible and, therefore, the maximum CP gain is $\gamma_I/\gamma_S$, where $\gamma_I$ and $\gamma_S$ are, respectively, gyromagnetic ratios for the abundant and rare spins.

Spin polarization is proportional to both the inverse spin temperature and the amplitude of the field. One may hope that even at equal spin temperatures it is possible reaching higher polarization of rare spins by a relative increase of the RF field amplitude. Efficient thermal contact exists only in a range of mismatches $|\omega_{1I} - \omega_{1S}| \sim \omega_{loc}$, where $\omega_{1I}$ and $\omega_{1S}$ are the RF fields amplitudes, and $\omega_{loc}$ is the amplitude of the dipolar interactions between the abundant spins (all amplitudes are in frequency units). The desired ratio $\omega_{1S}/\omega_{1I} \gg 1$ can be achieved only at $\omega_{1I} \ll \omega_{loc}$. Then, however, the spin order of the abundant nuclei would "leak" from the Zeeman to the dipolar reservoir. This brings the idea that, maybe, it is beneficial to recover magnetization from the dipolar reservoir, using the latter as an intermediate "storage". At $\omega_{1S}/\omega_{loc} > 1$ one may still expect some contact between the rare nuclei's Zeeman and the dipolar reservoirs, and an additional $\omega_{1S}/\omega_{loc}$ enhancement of the rare spins polarization.

Adiabatic "total cross-polarization" scheme[2-5] implements such two-step transfer of polarization. Several adiabatic CP schemes have been proposed in Ref.[2], and the mechanisms of adiabatic transfer have been further studied theoretically in Refs.[6,7] The "total cross-polarization" uses adiabatic demagnetization - remagnetization in the rotating frame (ADRF/ARRF), when the amplitude of RF field on abundant nuclei gradually decreases, and the polarization, initially locked by this field, is converted into the dipolar order. After that, the amplitude of RF field on rare nuclei increases, and the dipolar order is converted into polarization of rare nuclei.

This approach aims at taking advantage of an additional adiabatic amplification factor $(N_I/N_S)^{1/2}$, where $N_I$ and $N_S$ are, respectively, the numbers of abundant and rare spins in the sample. The method is conceptually different from various adiabatic modifications of the conventional Hartmann-Hahn cross polarization (HHCP), based on direct thermal contact between Zeeman reservoirs. The adiabatic passage through the Hartmann-Hahn condition[8-10] gives only a moderate increase of polarization, which remains below $\gamma_I/\gamma_S$. The review of such techniques can be found in Ref.[11] Their major advantage is that they are designed for and can be used under fast magic-angle spinning (MAS) of a sample.

Modern NMR spectrometers allow using shaped pulses with up to $10^5$ steps of individually set amplitudes and phases. This is sufficient to program long frequency-sweeping pulses with broad sweeping range. Such pulses can be conveniently applied for performing adiabatic demagnetization-remagnetization in the laboratory frame (ADLF/ARLF). The reference to the laboratory frame here will mean that the Zeeman reservoirs are formed by the spins' z-components, rather than the transverse components locked by the RF fields. Therefore, one can implement a two-step ACP using low-amplitude long frequency-sweeping pulses where, similar to the "total cross-polarization",[2-5] the Zeeman order of the abundant spin system is first converted into the dipolar order and then, into the Zeeman order of rare nuclei. The difference is that these processes are now performed in the laboratory frame. The major advantages of using ADLF/ARLF are that it does not require long irradiation with high RF power and that the relaxation times in the laboratory frame are longer. Therefore, one may expect that the process can be performed in a "softer" way, and better adiabaticity can be achieved.

**SPIN THERMODYNAMICS**

The principles of cross polarization are well described in the literature. The details can be found in the monographs.[12-15] Here we will only present some essential theoretical concepts necessary to understand how the ADLF/ARLF CP works. Let $\rho_{in}$ and $\rho_{fin}$ be the initial and final density matrices. In the initial state, the abundant nuclei $I$ are polarized, and in the final state, the rare nuclei $S$ are polarized. In the high-temperature approximation, the initial density matrix is

$$\rho_{in} = (1/2^N) (\mathbf{1} + \beta_{in} \omega_{0I} I_Z \otimes \mathbf{1}_S), \qquad (1)$$

and, similarly,

$$\rho_{fin} = (1/2^N) (\mathbf{1} + \beta_{fin} \omega_{0S} S_Z \otimes \mathbf{1}_I), \qquad (2)$$

where $N = N_I + N_S$ is the total number of spins, $I_Z$ and $S_Z$ are, respectively, total z-components for spins $I$ and $S$, $\omega_{0I}$ and $\omega_{0S}$ are their Larmor frequencies, $\beta = 1/kT$ is the inverse spin temperature, and $k$ is the Boltzmann constant. In the above equations, we explicitly used the identity operators for the two types of spins to emphasize that all operators are considered in the entire Hilbert space. If the process of converting the initial state into the final state is adiabatic, the entropy is conserved:

$$-k\, Tr(\rho_{in}\, ln\, \rho_{in}) = -k\, Tr(\rho_{fin}\, ln\, \rho_{fin}), \qquad (3)$$

or

$$\beta_{in}\, \omega_{0I}\, Tr(\rho_{in}\, I_Z \otimes \mathbf{1}_S) = \beta_{fin}\, \omega_{0S}\, Tr(\rho_{fin}\, S_Z \otimes \mathbf{1}_I), \qquad (4)$$

where the same high-temperature approximation has been used to simplify natural logs, and $Tr$ denotes trace of the matrix. By comparing this equation to the polarizations $P_I$ and $P_S$:

$$P_I = (2/N_I)\langle I_Z \rangle = (2/N_I)\, Tr(\rho_{in}\, I_Z \otimes \mathbf{1}_S), \qquad (5)$$

and

$$P_S = (2/N_S)\langle S_Z \rangle = (2/N_S)\, Tr(\rho_{fin}\, S_Z \otimes \mathbf{1}_I), \qquad (6)$$

one concludes that

$$\beta_{in}\, \omega_{0I}\, P_I\, N_I = \beta_{fin}\, \omega_{0S}\, P_S\, N_S, \qquad \text{or} \qquad P_I^2\, N_I = P_S^2\, N_S. \qquad (7)$$

Therefore, the polarization of rare spins can be increased by the additional adiabatic enhancement factor $P_S/P_I = (N_I/N_S)^{1/2}$. This ideal value can be reached only when the process is infinitely slow. In practice, $T_{1d}$ relaxation sets the time frame. The first step, conversion of the Zeeman order of abundant spins into the dipolar order can be very efficient, since the order per spin (entropy per spin) does not change. The "bottleneck" is the second step of converting the dipolar energy into the Zeeman energy of rare spins.

This process is slow and requires transfer of spin order from large distances when the concentration of rare nuclei is small. To estimate the actual performance of the ADLF/ARLF CP scheme one needs a description of the kinetics of this limiting step.

Proper kinetic equations can be obtained by applying the Provotorov's thermodynamic theory.[16,12] The Hamiltonian of our problem is

$$H = -\omega_{0S} S_Z + H_d + 2\omega_{1S} S_X \cos \omega t, \quad (8)$$

where $H_d$ is the Hamiltonian of dipole-dipole interactions. When $\omega_{1S}$ is small, the density matrix at all times has the quasi-equilibrium form

$$\rho(t) = (1/2^N) [1 + \beta_S(t) \omega_{0S} S_Z - \beta_d(t) H_d], \quad (9)$$

where $\beta_S(t)$ and $\beta_d(t)$ are, respectively, the Zeeman and dipolar inverse spin temperatures. In the initial state, $\beta_d(t) = \beta_d(0)$ and $\beta_S(t) = 0$. The kinetic equations for $\beta_S(t)$ and $\beta_d(t)$ can be derived by using the standard approach.[16,12] The result is

$$d\beta_S(t)/dt = - W[\beta_S(t) - (\Delta/\omega_{0S}) \beta_d(t)], \quad (10a)$$

$$d\beta_d(t)/dt = W \gamma [\beta_S(t) - (\Delta/\omega_{0S}) \beta_d(t)], \quad (10b)$$

where $W = \pi \omega_1^2 g(\Delta)$, $\Delta(t) = \omega_{0S} - \omega(t)$ is slowly changing resonance offset for rare nuclei, $g(\Delta)$ is their absorption line shape, and $\gamma = \omega_{0S}\Delta \, Tr(S_Z^2)/Tr(H_d^2)$. After estimating the traces at $N_I \gg N_S$, $\gamma = (\omega_{0S}\Delta/\omega_{loc}^2)(N_S/N_I)$, where $\omega_{loc}$ is the dipolar local field[12] for abundant nuclei. Solution of Eqs. (10), under the conditions that the change of $\Delta$ from zero to infinity is infinitely slow and that $g(\Delta)$ does not turn to zero, results in the ideal enhancement factor $(N_I/N_S)^{1/2}$. By analyzing the system of kinetic equations (10), one finds that enhancement on the order of $(N_I/N_S)^{1/2}$ is approached at $\Delta \sim \omega_{loc}(N_I/N_S)^{1/2}$. In practically interesting cases, $g(\Delta)$ may be too small at such offsets and, therefore, the adiabatic enhancement will be kinetically limited to much smaller values. One can also notice that broader dipolar line $g(\Delta)$ for rare nuclei provides better enhancement. As a result, rare nuclei, strongly coupled to many abundant nuclei, are expected to have higher enhancement factors. The described ACP scheme is still too far from practically reaching the ideal adiabatic enhancement. However, the experimental examples below show that it can provide a significant, 2 – 3 times, adiabatic enhancement for transferring polarization from protons to rare $^{13}$C nuclei.

**EXPERIMENTAL RESULTS**

The pulse sequence[17] is shown in Fig. 1. The first adiabatic frequency-sweeping pulse of low RF amplitude starts irradiation on the abundant nuclei far off-resonance and then, its frequency gradually approaches the center of the spectrum. In this process of adiabatic demagnetization, the Zeeman order of the abundant spins is converted into the dipolar order. What is important here is that the reservoir of dipole-dipole interactions is common for both types of nuclei. Even though the secular dipole-dipole interactions between spins with different gyromagnetic ratios consist of zz-terms only, a common dipolar temperature is established.[12] Rare nuclei are not irradiated at this point and do not change their projections on the z-axis. However, flip-flops between the abundant spins adjust orientations of the abundant spins in the vicinity of a rare nucleus according to the existing z-fields produced by this rare nucleus. The second adiabatic frequency-sweeping pulse is applied to the rare spins. Irradiation starts at the center of the spectrum of rare nuclei and then the offset gradually increases. This pulse converts the dipole-dipole order of the entire system into the Zeeman order of rare spins. The kinetics of this process is described in the previous section.

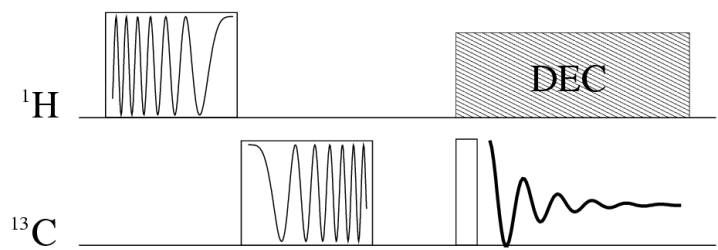

Fig. 1. NMR pulse sequence for adiabatic cross-polarization.

To test this scheme, we have chosen static samples of powdered solids, liquid crystal, and molecules dissolved in a liquid-crystalline matrix. The experiments have been performed using a Varian Unity/Inova 500 MHz NMR spectrometer. Both $^1$H and $^{13}$C adiabatic frequency-sweeping pulses were shaped pulses with constant RF amplitude and time-dependent phase. The number of constant-phase steps in the pulses ranged from 30 K to 50 K. The phase modulation was programmed as $f \times \tau_p \times \pi\,[2(t/\tau_p) - (t/\tau_p)^2]$, where $f$ is the frequency in Hz, $\tau_p$ is the duration of the frequency-sweeping pulse in sec, and $0 < t < \tau_p$. With this phase modulation, the pulse changes its frequency linearly with time from $f$ Hz to 0 Hz. The actual frequency at 0 Hz is the transmitter frequency, set by the spectrometer. SPINAL sequence[18] has been used for the heteronuclear decoupling.

**Solids**

The result for static powdered adamantane[17] is shown in Fig. 2. Compared to a single-pulse acquisition (Fig. 2a), the ACP scheme produced a six-fold increase in $^{13}$C polarization (Fig. 2b). In this experiment, the proton and carbon shaped pulses had, respectively, 100 ms and 300 ms durations, 40 kHz and 13 kHz sweeping ranges, and 1.0 kHz and 4.8 kHz RF fields' amplitudes ($\gamma B_1/2\pi$).

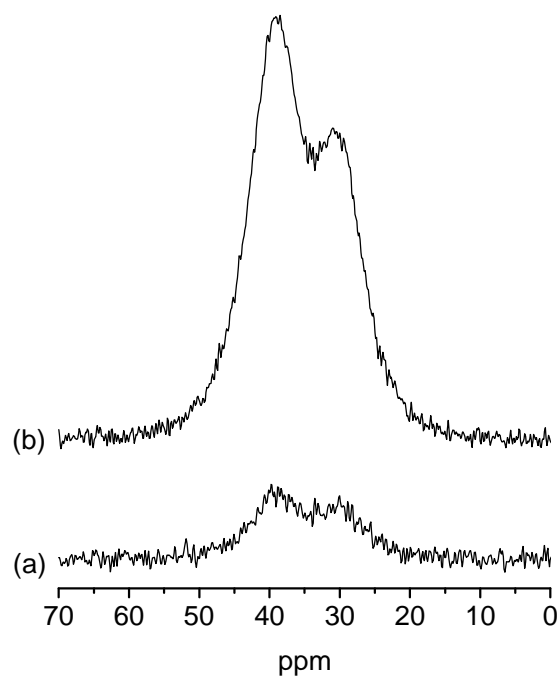

Fig. 2. $^{13}$C NMR spectra of adamantane at 298 K: (a) single-pulse acquisition; (b) adiabatic cross-polarization.

The next example of a static powdered solid is α-crystalline form of glycine. Isotopically labeled glycine-[$^{13}C_\alpha$,$^{15}$N] (ISOTEC) was diluted to 2% in natural abundance (n. a.) glycine (Aldrich), and the sample was recrystallized from aqueous solution. We have reached about seven-fold enhancement of $^{13}$C polarization by using the $^1$H adiabatic pulse of 100 ms duration, 100 kHz frequency-sweeping range, and 1.2 kHz amplitude ($\gamma B_1/2\pi$); the $^{13}$C pulse had 100 ms duration, 60 kHz frequency-sweeping range, and 9.5 kHz amplitude. The repetition time has been chosen long enough (25 s) to make the results independent of this time.

ADLF/ARLF performed by two consecutive frequency-sweeping pulses requires long time compared to other CP schemes. During this time, there is unavoidable loss of spin order due to spin-lattice relaxation of the dipolar reservoir ($T_{1d}$ process). Therefore, it is desirable to shorten the CP time. Another advantage of faster CP would be decreased experimental time. It is clear that the two adiabatic pulses can overlap in time, to some extent. When the first pulse is still supplying spin order to the dipolar reservoir, the second pulse can start converting this order into the Zeeman order of the rare spins. Experimentally, we have found that the adiabatic CP works well even when the two frequency-sweeping pulses are applied simultaneously. The achieved performance was, in most cases, even slightly better than with the consecutive pulses.

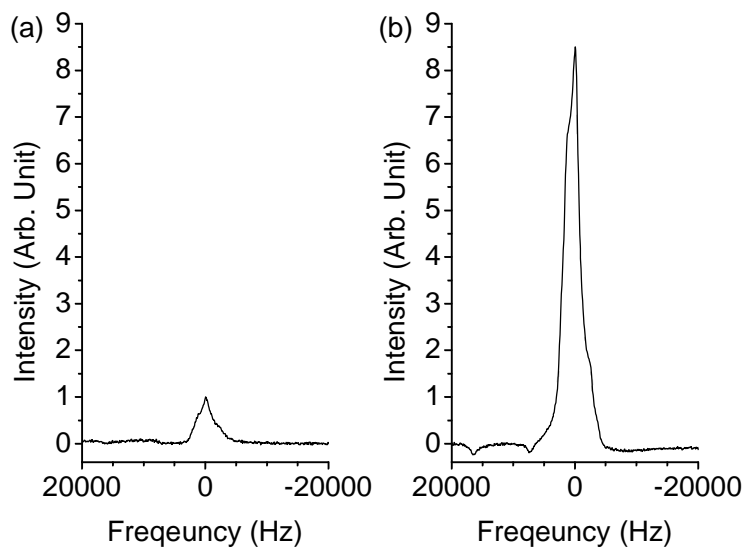

Fig. 3. $^{13}$C NMR spectra of 2% glycine-[$^{13}$C$_\alpha$,$^{15}$N] in n. a. glycine at 298 K: (a) single-pulse acquisition; (b) adiabatic cross-polarization with simultaneous frequency-sweeping pulses.

Fig. 3 shows the result of applying simultaneous frequency-sweeping pulses to the 2% glycine sample described above. Both pulses had 100 ms duration. The $^1$H adiabatic pulse had 100 kHz frequency-sweeping range and 3 kHz amplitude ($\gamma B_1/2\pi$); the $^{13}$C pulse had 100 kHz frequency-sweeping range and 9 kHz amplitude. As one can see in Fig. 3, the scheme with simultaneous pulses produced 8.5 times increase in the $^{13}$C polarization.

**Liquid crystal**

Liquid crystal 5CB has been used as purchased from Aldrich, without further purification. The results are shown in Fig. 4. The spectrum (a) is acquired with a single $^{13}$C pulse, starting with a thermal equilibrium state. The spectrum (b) is recorded after 2.65 ms CP with the matched Hartmann-Hahn RF fields of 19 kHz amplitudes. The result[17] of the ACP with the pulse sequence in Fig. 1 is shown in Fig. 4c. Each of the two frequency-sweeping pulses had 100 ms duration and 40 kHz frequency sweeping range. The RF fields' amplitudes ($\gamma B_1/2\pi$) for $^1$H and $^{13}$C pulses were 1.6 kHz and 4.2 kHz, respectively. The spin-lattice relaxation time of the dipole-dipole reservoir $T_{1d}$ was measured to be 0.54 s. The numbers near the peaks in Figs. 4b and 4c show the relative intensities of those peaks compared to the single-pulse acquisition. One can see that the $^{13}$C magnetization created by our ACP sequence is significantly higher than the magnetization resulting from the conventional HHCP.

The scheme with simultaneous pulses showed very similar performance for 5CB. The result is shown in Fig. 5. Here we used the same frequency-sweeping pulses as in Fig. 4 and optimized the reference frequencies of the adiabatic pulses to maximize the signal in the aliphatic region of the $^{13}$C spectrum.

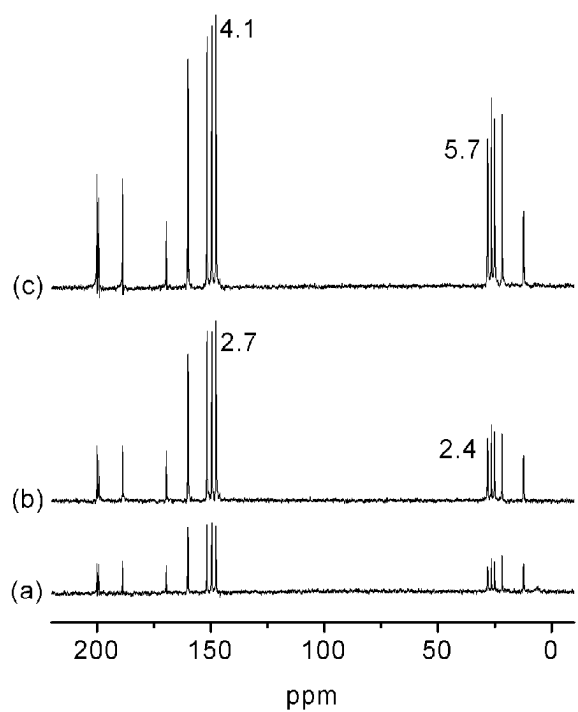

Fig. 4. $^{13}$C NMR spectra of 5CB at 298 K: (a) single-pulse acquisition; (b) HHCP; (c) adiabatic cross-polarization with consecutive pulses.

**Molecules in liquid-crystalline solvent**

Anisotropic liquid-crystalline solvents are routinely used for small to moderate-size molecules to retrieve structural information from the residual dipole-dipole couplings. The residual couplings are usually much weaker than the couplings in solids or liquid crystals. Such systems are studied with liquid-state NMR techniques, and to enhance the $^{13}$C polarization one would probably use the nuclear Overhauser effect (NOE).

We have tested the scheme with simultaneous frequency-sweeping pulses for molecules dissolved in a liquid-crystalline matrix. The sample was 2% 1-dodecene-1,2-$^{13}$C$_2$ (Aldrich) in 5CB. The same frequency-sweeping pulses, as used for 5CB, have been applied. The frequencies and RF amplitudes of the pulses have been optimized for maximizing the signal from the solute molecule. The RF amplitudes were about 4 kHz for both the $^1$H and $^{13}$C pulses. The result in Fig. 6 shows that the ACP increases the peaks heights by 3.5 – 4.4 times. Typically, NOE can increase $^{13}$C signal in liquids by about 2 times, and the theoretical maximum is 3 in the limit of rapid molecular motion.[15] Therefore, for molecules partially oriented in liquid-crystalline solvents, ACP can perform better than either NOE or conventional CP.

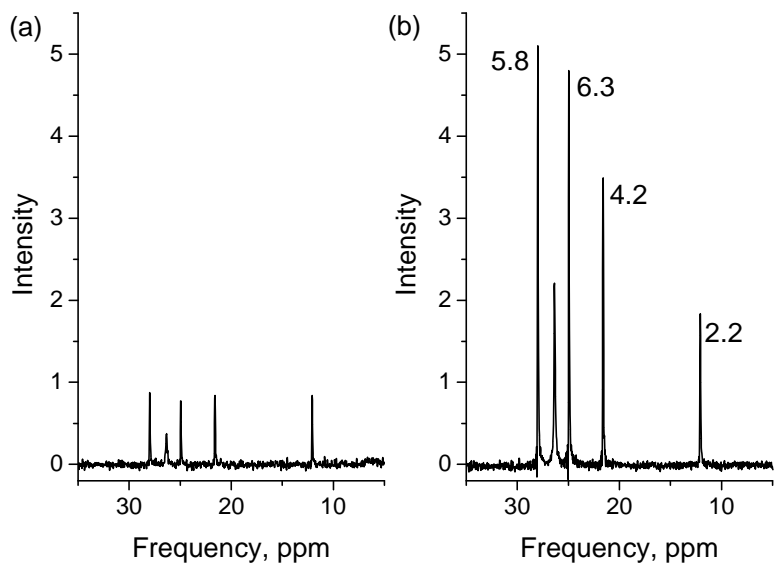

Fig. 5. $^{13}$C NMR spectra of 5CB at 298 K: (a) single-pulse acquisition; (b) adiabatic cross-polarization with simultaneous frequency-sweeping pulses.

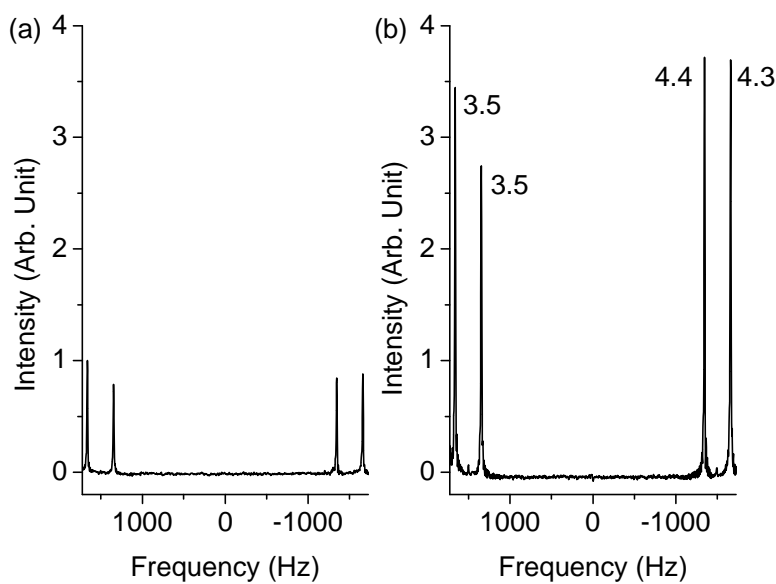

Fig. 6. $^{13}$C NMR spectra of 2% 1-dodecene-1,2-$^{13}$C$_2$ in 5CB at 298 K: (a) single-pulse acquisition; (b) adiabatic cross-polarization with simultaneous frequency-sweeping pulses.

## CONCLUSION

In conclusion, modern NMR spectrometers can conveniently implement CP via adiabatic demagnetization-remagnetization in the laboratory frame, by applying low-power shaped frequency-sweeping pulses. In the described ACP technique, the dipolar reservoir is used as an intermediate storage of spin order and, therefore, the method is limited to static samples. This is a serious limitation since many solid-state NMR experiments require magic-angle spinning of a sample. At the same time, for static solids, liquid crystals, or molecules in liquid-crystalline solvents, the described version of ACP is absolutely superior to any existing cross-polarization techniques.